\documentclass[a4paper,11pt]{article}
\usepackage{pos}
\usepackage{amsmath}
\usepackage{epsfig}

\let\oldsqrt\sqrt
\def\sqrt{\mathpalette\DHLhksqrt}
\def\DHLhksqrt#1#2{%
\setbox0=\hbox{$#1\oldsqrt{#2\,}$}\dimen0=\ht0
\advance\dimen0-0.2\ht0
\setbox2=\hbox{\vrule height\ht0 depth -\dimen0}%
{\box0\lower0.4pt\box2}}

\newcommand{\sss}[1]{{\scriptscriptstyle{#1}}}

\newcommand{\uPl}{\mathrm{Pl}}

\newcommand{\usssPl}{\sss{\uPl}}

\newcommand{\Mp}{M_\usssPl}

\newcommand{\beq}{\begin{equation}}
\newcommand{\eeq}{\end{equation}}
\newcommand{\bea}{\begin{equation}\begin{aligned}}
\newcommand{\eea}{\end{aligned}\end{equation}}

\newlength{\wsingfig}
\newlength{\wdblefig}
\newlength{\wquadfig}
\newlength{\wtriplefig}
\newcommand{\Eq}[1]{Eq.~(\ref{#1})}

\newcommand{\Sec}[1]{Sec.~\ref{#1}}

\title{The $H_0$ tension alleviated through ultra-light primordial black holes: an information insight through gravitational waves}

\author*[a]{Theodoros Papanikolaou}

\affiliation[a]{National Observatory of Athens, Lofos Nymfon, 11852 Athens, 
Greece}

\emailAdd{papaniko@noa.gr}

\abstract{The Hawking evaporation of ultra-light primordial black holes (PBH) dominating the early Universe before Big Bang Nucleosynthesis can potentially increase the effective number of extra neutrino species $\Delta N_\mathrm{eff}$ through the emission of dark radiation degrees of freedom alleviating in this way the $H_0$ tension problem. Interestingly, these light PBHs can form a gas of Poisson distributed compact objects which can induce a gravitational-wave (GW) background due to second order gravitational interactions. Therefore, by considering the contribution to $\Delta N_\mathrm{eff}$ due to the production of the aforementioned GW background we revisit in this work the constraints on the relevant parameters at hand, namely the PBH mass,  $m_\mathrm{PBH}$, the initial PBH abundance at PBH formation time, $\Omega_\mathrm{PBH,f}$ and the number of DR radiation degrees of freedom, $g_\mathrm{DR}$ by accounting at the same time for the relevant upper bounds constraints on $\Delta N_\mathrm{eff}$ from the Planck collaboration.

}

\FullConference{%
  Corfu Summer Institute 2022 "School and Workshops on Elementary Particle Physics and Gravity",\\
  28 August - 1 October, 2022\\
  Corfu, Greece}

\begin{document}
\maketitle

\section{Introduction}
Primordial black holes (PBHs),  introduced in the early `70s~\cite{1967SvA....10..602Z, Carr:1974nx} are currently attracting an increasing attention since they can solve in a natural way a plethora of issues of modern cosmology.  In particular, they can potentially account for a part or the totality of the dark matter~\cite{Chapline:1975ojl}, explain the generation of large-scale structures (LSS) through Poisson fluctuations~\cite{Meszaros:1975ef} and seed the supermassive black holes residing in galactic centers~\cite{1984MNRAS.206..315C}. Furthermore,  they can constitute viable candidates for the progenitors of the black-hole merging events recently detected by the LIGO/VIRGO collaboration~\cite{LIGOScientific:2018mvr} being associated with numerous gravitational-wave (GW) signals such GWs from PBH mergers~\cite{Nakamura:1997sm, Ioka:1998nz} and GWs induced by primordial curvature perturbations~\cite{Bugaev:2009zh, Saito_2009} [See here~\cite{Domenech:2021ztg} for a review].

Interestingly, ultra-light PBHs with masses $m_\mathrm{PBH}<10^9\mathrm{g}$ evaporating before Big Bang Nucleosynthesis (BBN)~\cite{Kawasaki:1999na,Kawasaki:2000en,Hasegawa:2019jsa,Carr:2020gox} are currently poorly constrained and are associated with a very interesting phenomenology.  In particular, these ultra-light PBHs can drive early PBH-matter dominated eras~\cite{Inomata:2019zqy,Papanikolaou:2020qtd,Domenech:2020ssp} before BBN producing at the same time the DM relic abundance and the hot Standard Model (SM) plasma~\cite{Lennon:2017tqq} reheating at the same time the Universe through their evaporation~\cite{Martin:2019nuw}. Furthermore, they can potentially alleviate the Hubble tension~\cite{Hooper:2019gtx,Nesseris:2019fwr,Lunardini:2019zob} and produce as well naturally the baryon assymetry through CP violating out-of-equilibrium decays of their Hawking evaporation products~\cite{Barrow:1990he,Bhaumik:2022pil,Bhaumik:2022zdd,Gehrman:2022imk}.  Regarding their production mechanism, these ultra-light PBHs can be abundantly produced within inflationary setups~\cite{Martin:2019nuw,Briaud:2023eae} as well as within quantum gravity~\cite{Papanikolaou:2023crz} and bouncing cosmological setups~\cite{Banerjee:2022xft}.

In this work, we revisit the scenario of the Hawking evaporation of ultra-light PBHs efficiently accounting for the alleviation of the Hubble tension through the injection to the primordial plasma of dark radiation (DR) light degrees of freedom with feeble couplings to the Standard Model (SM).  Interestingly, as it was shown in~\cite{Hooper:2019gtx,Nesseris:2019fwr,Lunardini:2019zob} the production of DR degrees of freedom can potentially increase the effective number of extra neutrino species $\Delta N_\mathrm{eff}$ and subsequently the value of the Hubble parameter at early times~\cite{Planck:2018vyg} reconciling it in this way with its late-time value as measured by late-time observational probes~\cite{Riess:2020fzl}.  In particular, as it is proposed in~\cite{Riess:2016jrr,Riess:2018byc,Riess:2019cxk} a value of $\Delta N_\mathrm{eff}\sim 0.1 - 0.3$ would be enough to substantially relax the $H_0$ tension. To gain then an insight on this PBH motivated alleviation mechanism of the Hubble tension, we will make use  of the portal of GWs induced by the gravitational potential of a population of randomnly distributed PBHs~\cite{Papanikolaou:2020qtd,Domenech:2020ssp}.

The paper is organised as follows: In \Sec{sec:BH_Evaporation} we recast the basics of black hole evaporation while in \Sec{sec:PBH_domination} we derive the necessary conditions for an early PBH-dominated Universe. Then, in \Sec{sec:DR_Delta_N_eff} we extract the contribution of DR light degrees of freedom to $\Delta N_\mathrm{eff}$ while in \Sec{sec:SIGW_Poisson_PBH} after reviewing briefly the GWs associated to Poisson PBH energy density fluctuations we extract their contribution to $\Delta N_\mathrm{eff}$. Followingly, in \Sec{sec:revisit_early_PBH_domination} by accounting for the contribution to $\Delta N_\mathrm{eff}$ from both dark radiation and GWs we revisit the constraints on the relevant parameters at hand, namely the PBH mass,  $m_\mathrm{PBH}$, the initial PBH abundance at PBH formation time, $\Omega_\mathrm{PBH,f}$ and the number of DR radiation degrees of freedom, $g_\mathrm{DR}$. Finally, \Sec{sec:conclusions} is devoted to conclusions.

\section{The basics of black hole evaporation}\label{sec:BH_Evaporation}
Black holes radiate energy by emitting particles through the process of Hawking evaporation~\cite{Hawking:1975vcx} with their mass loss rate being recast as
\beq\label{eq:BH_mass_loss_rate}
\frac{\mathrm{d}m_\mathrm{BH}}{\mathrm{d}t}= -\frac{\mathcal{G}g_\mathrm{*,H}(T_\mathrm{BH})\Mp^4}{30720\pi m^2_\mathrm{BH}}, 
\eeq
with $\mathcal{G}\simeq 3.8$ being the appropriate grey factor for Schwarzschild BHs and $T_\mathrm{BH}$ being the BH temperature which can be recast as
\beq
T_\mathrm{BH}\equiv \frac{M^2_\mathrm{Pl}}{8\pi m_\mathrm{BH}} \simeq 1.05\times 10^{13}\mathrm{GeV}\left(\frac{\mathrm{g}}{m_\mathrm{PBH}}\right).
\eeq
The factor $g_\mathrm{*,H}(T_\mathrm{BH})$ counts all the existing degrees of freedom with mass $m$ below $T_\mathrm{BH}$, i.e. $m<T_\mathrm{BH}$ according to the prescription~\cite{MacGibbon:1990zk,MacGibbon:1991tj}

\beq
g_\mathrm{*,H}(T_\mathrm{BH}) = \sum_i w_ig_{i,\mathrm{H}}, \quad g_{i,\mathrm{H}} = \begin{cases} 1.82, \quad s=0\\
1.0 \quad s=1/2\\ 0.41 \quad s = 1\\ 0.05\quad s = 2,
\end{cases}
\eeq
where $w_i = 2s_i +1$ for massive particle species with spin $s_i$ and $w_i = 2$ for massless particles with $s_i>0$. Since particles species with mass $m$ are Hawking radiated whenever $m<T_\mathrm{BH}$ according to the above prescription one can show that for temperatures higher than the electroweak scale$\sim 100\mathrm{GeV}$ practically all the SM particles are emitted whereas for temperatures below the $\mathrm{MeV}$ scale only photons and neutrinos are emitted.  Thus, one can approximately recast $g_\mathrm{*,H}(T_\mathrm{BH})$ as
\beq\label{eq:g_*H}
g_\mathrm{*,H}(T_\mathrm{BH}) \simeq \begin{cases} 108, \quad T_\mathrm{BH}\gg 100\mathrm{GeV}, \quad m_\mathrm{PBH}\ll 10^{11}\mathrm{g}
\\
7, \quad T_\mathrm{BH}\ll 1\mathrm{MeV}, \quad m_\mathrm{PBH}\gg 10^{16}\mathrm{g}.
\end{cases}
\eeq
Assuming therefore that $g_\mathrm{*,H}(T_\mathrm{BH})$ is constant for BHs evaporating before BBN, which is always true as one can see from \Eq{eq:g_*H} one can solve \Eq{eq:BH_mass_loss_rate} finding at the end that the time needed for a BH to complete its evaporation  reads as
\beq\label{eq:Dt_evap}
\Delta t_\mathrm{evap} = \frac{160}{\pi g_\mathrm{*,H}(T_\mathrm{BH})} \frac{m^3_\mathrm{PBH}}{\Mp^4}.
\eeq
One would also consider that  BHs can potentially increase their mass through the process of mergers and accretion. Regarding the effect of accretion, recent analyses~\cite{DeLuca:2020bjf,DeLuca:2020fpg} showed that within the regime of Bondi-Hoyle type accretion~\cite{1944MNRAS.104..273B} accretion is negligible when $m_\mathrm{BH} < O(10)M_\odot$.  Concerning the effect of BH mergers this will be important only at very early times, corresponding to $T\succsim 10^8\mathrm{GeV}\times (10^8 \mathrm{g}/m_\mathrm{PBH})^{3/4}$ [See Appendix A of~\cite{Hooper:2019gtx}]. Thus, $m_\mathrm{PBH}$ should be regarded as the BH mass after the process of merging has stopped to be efficient.

\section{The primordial black hole dominated Universe}\label{sec:PBH_domination}
PBHs form standardly in the radiation-dominated (RD) era out of the collapse of enhanced cosmological perturbations. Given the fact that they behave as dust within a RD background their abundance $\Omega_\mathrm{PBH}$ will scale as 
\beq\label{eq:Omega_PBH}
\Omega_\mathrm{PBH} =\frac{ \rho_\mathrm{PBH}}{\rho_\mathrm{r}}\propto \frac{a^{-3}}{a^{-4}}\propto a.
\eeq

Thus, since $\Omega_\mathrm{PBH}\propto a$, at some point PBHs will dominate the energy budget of the Universe when $\Omega_\mathrm{PBH}=1$.  To compute therefore, the necessary conditions for a PBH dominated Universe we will assume monochromatic PBH mass distributions and require that PBHs evaporate after BBN time. Thus, since after BBN time the Universe continues to evolve in a RD dominated era up to matter-radiation equality at redshift $z_\mathrm{eq}\sim 1100$, we will require that PBHs dominate the energy Universe content before their evaporation.  In particular,  by requiring that  $\Omega_\mathrm{PBH}=1$, from \Eq{eq:Omega_PBH} we get that $a_\mathrm{d}=a_\mathrm{f}/\Omega_\mathrm{PBH,f}$ where the index $\mathrm{d}$ stands for the onset of the PBH-dominated era. At the end, accounting for the fact that during radiation domination era $H\simeq 1/(2t)$, and demanding  that $t_\mathrm{evap}>t_\mathrm{d}$, one obtains that in order to have an early matter-dominated (MD) era driven by PBHs, the PBH abundance at PBH formation time is bounded from below according to the following expression:
\bea
\label{eq:domain:OmegaPBHf}
\Omega_\mathrm{PBH,f} > 10^{-15} \sqrt{\frac{g_\mathrm{*,H}(T_\mathrm{BH})}{108}} \frac{10^9\mathrm{g}}{m_\mathrm{PBH}}.
\eea
For the above expression, we used as well the fact that the Hawking evaporation time of a black hole scales with $m_\mathrm{PBH}$ like in \Eq{eq:Dt_evap}. Since the PBHs we consider here form before BBN $g_\mathrm{*,H}(T_\mathrm{BH})\simeq 108$ as dictated by the prescription \ref{eq:g_*H}.

Regarding the PBH mass range we consider PBHs forming after the end of inflation and evaporate before BBN time. Thus, one can derive a lower and an upper bound on $m_\mathrm{PBH}$ by accounting for the current Planck upper bound on the tensor-to-scalar ratio for single-field slow-roll models of inflation, which gives  $\rho^{1/4}_\mathrm{inf}<10^{16}\mathrm{GeV}$ ~\cite{Planck:2018jri} as well a conservative lower bound on the reheating energy scale, i.e. $\rho^{1/4}_\mathrm{reh}> 4\mathrm{MeV}$~\cite{Kawasaki:1999na,Kawasaki:2000en,Hasegawa:2019jsa,Carr:2020gox}. Finally, by requiring that $\rho_\mathrm{reh}\geq\rho_\mathrm{BBN}$ and accounting for the fact that the PBH mass is roughly equal to the mass inside the cosmological horizon at PBH fomation time,  which is standardly considered as the horizon crossing time of the collapsing scale during the RD era, one can show that the relevant PBH mass range is given by 
\bea
\label{eq:domain:mPBHf}
10 \mathrm{g}< m_\mathrm{PBH}< 10^{9} \mathrm{g}\,.
\eea

\section{The contribution of dark radiation to $\Delta N_\mathrm{eff}$}\label{sec:DR_Delta_N_eff}
After the end of PBH evaporation, the primordial plasma is filled with SM radiation alongside with any other Hawking radiated product of any hidden matter sector. If however within the Hawking emitted particles of the hidden sector there exist light degrees of freedom with feeble couplings (perhaps only gravitational) to the SM particles, these products, usually called ``dark radiation" (DR), will not thermalise and will contribute to the radiation energy content of the Universe.  A way to test there existence is by measuring their contribution to the effective number of neutrino species, $N_\mathrm{eff}$ defined as $N_\mathrm{eff}\equiv N^\mathrm{SM}_\nu + \Delta N_\mathrm{eff}$, where $N^\mathrm{SM}_\nu$ is the effective number of the SM neutrino species and $\Delta N_\mathrm{eff}$ is the effective number of extra neutrino species.

Let us now derive the contribution to $\Delta N_\mathrm{eff}$ from DR degrees of freedom. To do so, one should write the total radiation energy density as the sum of the energy densities of fermions and bosons in the primordial thermal bath, namely as
\beq\label{eq:rho_rad}
\rho_\mathrm{rad} = \frac{\pi^2}{30}\left[\sum_\mathrm{B} g_{\mathrm{B}}\left(\frac{T_\mathrm{B}}{T}\right)^4  + \frac{7}{8}\sum_\mathrm{F}g_{\mathrm{F}}\left(\frac{T_\mathrm{F}}{T}\right)^4\right]T^4,
\eeq
where the sum is performed over all the fermionic ($g_\mathrm{F}$) and bosonic ($g_\mathrm{B}$) degrees of freedom that are relativistic, namely whose mass $m_i$ is smaller than the temperature of the thermal bath, i.e. $T>m_\mathrm{i}$. One can also define the effective number of energy density relativistic degrees of freedom defined as 
\beq\label{eq:g_*}
g_{*} (T) \equiv \sum_\mathrm{B} g_{\mathrm{B}}\left(\frac{T_\mathrm{B}}{T}\right)^4  + \frac{7}{8}\sum_\mathrm{F}g_{\mathrm{F}}\left(\frac{T_\mathrm{F}}{T}\right)^4.
\eeq

After neutrino decoupling at around $1\mathrm{MeV}$, the electrons become non relativistic at $T\simeq 500keV$ and one is met with lots of electron-positron annihilation process which produces an excess of photons.  Since neutrinos have already decoupled from the thermal bath, the entropy of electrons and positrons is transferred mainly to photons. Thus, one is met with a slightly smaller temperatures of neutrinos $T_\mathrm{\nu}$ compared to that of photons $T_\mathrm{\gamma}$ which reads as
\beq\label{eq:T_nu_vs_T_gamma}
T_\mathrm{\nu} = \left(\frac{4}{11}\right)^{1/3}T_\mathrm{\gamma}.
\eeq
At the end,  after the epoch of electron-proton annihilation the only relativistic species in the primordial thermal bath are the photons and the neutrinos. One then can infer from \Eq{eq:rho_rad} and \Eq{eq:T_nu_vs_T_gamma} that the radiation energy density after the epoch of $e^{+}e^{-}$ annihilation is related to the photon energy density as follows:

\beq\label{eq:rho_rad_vs_rho_gamma}
\rho_\mathrm{rad}=\rho_\mathrm{\gamma}\left[1+\frac{7}{8}\left(\frac{4}{11}\right)^{4/3}N_\mathrm{eff}\right],
\eeq
where  $N_\mathrm{eff}$ is the effective number of neutrino species defined above.

Assuming the presence of extra light dark radiation degrees of freedom who have decoupled from the thermal bath given their feeble coupling with SM particles, one can write the total radiation energy density as the sum of the SM radiation density plus the DR radiation density, namely $\rho_\mathrm{rad} = \rho^\mathrm{SM}_\mathrm{rad} + \rho_\mathrm{DR}$. Using therefore \Eq{eq:rho_rad_vs_rho_gamma} and writing $N_\mathrm{eff}$ as $N_\mathrm{eff}= N^\mathrm{SM}_\nu + \Delta N_\mathrm{eff}$ one can straightforwardly show that at matter-radiation equality ($t=t_\mathrm{eq}$) at $T=0.75\mathrm{eV}$ one gets that the effective number of extra neutrino species $ \Delta N_\mathrm{eff}$ due to dark radiation reads as
\beq\label{eq:DeltaN_eff_rho_DR}
\Delta N_\mathrm{eff} \rvert _\mathrm{DR}= \frac{\rho_\mathrm{DR}(t_\mathrm{eq})}{\rho^\mathrm{SM}_\mathrm{rad}(t_\mathrm{eq})} \left[\frac{8}{7}\left(\frac{11}{4}\right)^{4/3}+N^\mathrm{SM}_\nu\right],
\eeq
where $N^\mathrm{SM}_\nu=3.046$\footnote{Neutrino decoupling was not exactly quite complete when $e^{+}e^{-}$ annihilation began taking place. Thus,  some of the energy and entropy of electrons and positrons passed to neutrinos as well. Taking this into account the effective number of  SM neutrinos gets a bit enhanced above $3$, namely $N^\mathrm{SM}_\nu=3.046$.}. Then, accounting for the conservation of entropy one gets that $T\sim g^{-1/3}_{*,S}a^{-1}$ where $g_{*,S}$ is the effective number of the entropy relativistic degrees of freedom defined as
\beq\label{eq:g_S}
g_{*,S} (T)\equiv \sum_\mathrm{B} g_{\mathrm{B}}\left(\frac{T_\mathrm{B}}{T}\right)^3  + \frac{7}{8}\sum_\mathrm{F}g_{\mathrm{F}}\left(\frac{T_\mathrm{F}}{T}\right)^3.
\eeq
At the end, applying the entropy conservation between the PBH evaporation time and the matter-radiation equality time and accounting for the fact that $g_{*,S} (T) = g_{*} (T)$ up to neutrino decoupling one obtains that 
 \beq
 \frac{\rho_\mathrm{DR}(t_\mathrm{eq})}{\rho^\mathrm{SM}_\mathrm{rad}(t_\mathrm{eq})} = \frac{\rho_\mathrm{DR}(t_\mathrm{evap})}{\rho^\mathrm{SM}_\mathrm{rad}(t_\mathrm{evap})}\left(\frac{g_{*,\mathrm{S}}(T_\mathrm{eq})}{g_{*,\mathrm{S}}(T_\mathrm{evap})}\right)^{1/3}\frac{g_{*,S}(T_\mathrm{eq})}{g_{*}(T_\mathrm{eq})},
 \eeq
where the ratio $\frac{g_{*,S}(T_\mathrm{eq})}{g_{*}(T_\mathrm{eq})}$ will not be equal to one due to the fact that after the epoch of $e^{+}e^{-}$ annihilation entropy is transferred mainly to photons. In particular, one can show that $g_{*,S}(T_\mathrm{eq}) = 3.94$ and $g_{*}(T_\mathrm{eq}) = 3.36$. The ratio now $\rho_\mathrm{DR}(t_\mathrm{evap})/\rho^\mathrm{SM}_\mathrm{rad}(t_\mathrm{evap})$ will be equal to $g_{\mathrm{DR},H}/g_{*,H}$ given the fact that SM radiation and dark radiation are emitted from the thermal process of Hawking radiation with their degrees of freedom being counted at the end of the PBH evaporation by $g_{*,H}$ given by \Eq{eq:g_*H}.  Finally, one can recast $\Delta N_\mathrm{eff} \rvert _\mathrm{DR}$ as
\beq\label{eq:DeltaN_eff_DR}
\Delta N_\mathrm{eff} \rvert _\mathrm{DR} = 0.10 \frac{g_{\mathrm{DR},H}}{4}\left(\frac{106}{g_*(T_\mathrm{evap})}\right)^{1/3}.
\eeq
 
\section{The contribution of gravitational waves induced from Poisson primordial black hole fluctuations to $\Delta N_\mathrm{eff}$}\label{sec:SIGW_Poisson_PBH}
Up to now we have considered the contribution to $\Delta N_\mathrm{eff} $ from DR degrees of freedom with no significant coupling to the standard model. However, apart from DR degrees of freedom primordial gravitational waves (GWs) will contribute as well to $\Delta N_\mathrm{eff} $. In the following, we will consider GWs induced by the gravitational potential of a gas of randomnly distributed ultralight PBHs which can dominate the energy budget of the Universe before their evaporation taking place before BBN. In particular,  assuming that PBHs are randomly distributed at formation time (i.e. they have Poisson statistics),  their energy density is inhomogeneous while the total background energy density can be considered homogeneous.  Thus,  the energy density perturbation of the PBH matter field can be described by an isocurvature Poisson fluctuation.  Interestingly, as it was shown in~\cite{Papanikolaou:2020qtd},  this initial isocurvature PBH energy density perturbation in the radiation-dominated (RD) era, when PBHs are supposed to form, will convert into an adiabatic curvature perturbation deep in the PBH dominated era giving rise to the following power spectrum for the PBH gravitational potential $\Phi$:
\begin{equation}
\label{eq:PowerSpectrum:Phi:PBHdom}
\mathcal{P}_\Phi(k) = \frac{2}{3\pi} \left( \frac{k}{k_{\rm{UV}}} \right)^3 \left(5+\frac{4}{9}\frac{k^2}{k_{\rm{d}}^2} \right)^{-2}\, ,
\end{equation}
where $k_{\rm{d}}$ is the comoving scale exiting the Hubble radius at PBH domination time and $k_\mathrm{UV}$ stands for a UV-cutoff scale defined as $k_{\rm{UV}}\equiv a/\bar{r}$, where $\bar{r}$ corresponds to the mean PBH separation distance. Note that $k>k_{\mathrm{UV}}$ corresponds to distances within the mean separation distance, where the granularity of the PBH matter field and the associated non-linear effects become important.  For this reason, we restrict ourselves in the following to regions where $k<k_{\rm{UV}}$. 

Then, due to second order gravitational interactions [See~\cite{Domenech:2021ztg} for details], the above power spectrum associated to the PBH Poisson energy density fluctuations will induce GWs which can contribute to the effective number of extra neutrino species $\Delta N_\mathrm{eff} $.  The GWs induced by Poisson PBH fluctuations can be produced in three different phases: in the early RD era, in the PBH dominated era as well in the later RD era followed PBH evaporation. In particular,  the dominant contribution is due to the resonant GW production at $k\sim k_\mathrm{UV}$ right after the onset of the late RD era after PBH evaporation~\cite{Inomata:2019ivs}. Consequently, in the region close to the UV cutoff scale, $k\sim k_\mathrm{UV}$, where we expect the dominant GW production, one can approximate the GW spectrum today as follows~\cite{Domenech:2020ssp}:
\beq\label{eq:Omega_GW_t_0}
\Omega_\mathrm{GW}(t_0,k\sim k_\mathrm{UV}) \simeq 2\times 10^{40}\Omega^{16/3}_\mathrm{PBH,f}\left(\frac{m_\mathrm{PBH}}{10^9\mathrm{g}}\right)^{34/9}.
\eeq
At the end, one can write an analogous equation to \Eq{eq:DeltaN_eff_rho_DR} for the GW contribution to $\Delta N_\mathrm{eff} $, namely
\beq\label{eq:DeltaN_eff_rho_GW}
\Delta N_\mathrm{eff} \rvert _\mathrm{SIGW}= \frac{\rho_\mathrm{GW}(t_\mathrm{eq})}{\rho^\mathrm{SM}_\mathrm{rad}(t_\mathrm{eq})} \left[\frac{8}{7}\left(\frac{11}{4}\right)^{4/3}+N^\mathrm{SM}_\nu\right] =   \frac{\Omega_\mathrm{GW}(t_0)}{\Omega^\mathrm{SM}_\mathrm{rad}(t_\mathrm{0})} \left[\frac{8}{7}\left(\frac{11}{4}\right)^{4/3}+N^\mathrm{SM}_\nu\right],
\eeq
where for the last equality we accounted for the fact that both $\Omega_\mathrm{GW}$ and $\Omega^\mathrm{SM}_\mathrm{rad}$ scales as $a^{-4}$.  Accounting then for \Eq{eq:Omega_GW_t_0} as well for the fact that one gets that $\Omega_\mathrm{GW}(t_0)\simeq 10^{-5}$ one gets that 
\beq\label{eq:DeltaN_eff_Omega_GW}
\Delta N_\mathrm{eff} \rvert _\mathrm{SIGW}\simeq 6\times 10^{45}\Omega^{16/3}_\mathrm{f}\left(\frac{m_\mathrm{PBH}}{10^9\mathrm{g}}\right)^{34/9}.
\eeq

\section{The overall contribution to $\Delta N_\mathrm{eff}$}\label{sec:revisit_early_PBH_domination}
Combining \Eq{eq:DeltaN_eff_DR} and \Eq{eq:DeltaN_eff_Omega_GW} one gets that the overall contribution to  the effective number of extra neutrino species reads as
\beq
\Delta N_\mathrm{eff}  = \Delta N_\mathrm{eff} \rvert _\mathrm{DR} + \Delta N_\mathrm{eff} \rvert _\mathrm{SIGW} =  0.10 \frac{g_{\mathrm{DR},H}}{4}\left(\frac{106}{g_*(T_\mathrm{evap})}\right)^{1/3} + 6\times 10^{45}\Omega^{16/3}_\mathrm{f}\left(\frac{m_\mathrm{PBH}}{10^9\mathrm{g}}\right)^{34/9}.
\eeq
Let us now make an investigation of the dominant contribution to $\Delta N_\mathrm{eff}$ and extract constraints on the relevant parameters at hand.  

\subsection{$\Delta N_\mathrm{eff}\simeq\Delta N_\mathrm{eff} \rvert _\mathrm{DR}$ }

In particular, if the DR degrees of freedom emitted out of the process of Hawking evaporation is the dominant contribution to $\Delta N_\mathrm{eff}$ then one has that $ \Delta N_\mathrm{eff} \rvert _\mathrm{DR}\geq\Delta N_\mathrm{eff} \rvert _\mathrm{SIGW}$ leading to a upper bound constraint on the PBH mass, namely that
$$
m_\mathrm{PBH}\leq 30\Mp\left(\frac{g_\mathrm{DR,H}}{12}\right)^{9/34}\left(\frac{108}{g_*(T_\mathrm{evap})}\right)^{3/34}\Omega^{-24/17}_\mathrm{PBH,f}
$$
Then, by requiring that the above upper bound on the PBH mass is less than $10^9\mathrm{g}$ so as that we have PBH evaporation before BBN and requiring that PBH can drive for some time the Universe expansion before BBN [See \Eq{eq:domain:OmegaPBHf}] one can obtain a lower bound on the PBH mass, namely that 
$$m_\mathrm{PBH}\geq 500\mathrm{g}\left(\frac{12}{g_\mathrm{DR,H}}\right)^{3/16}\left(\frac{g_*(T_\mathrm{evap})}{108}\right)^{1/16}.$$
At the end, one gets that in order for the DR degrees of freedom to contribute dominantly to $\Delta N_\mathrm{eff}$ the PBH mass should lie within the following mass range
\beq
500\mathrm{g}\left(\frac{12}{g_\mathrm{DR,H}}\right)^{3/16}\left(\frac{g_*(T_\mathrm{evap})}{108}\right)^{1/16}\leq m_\mathrm{PBH}\leq 30\Mp\left(\frac{g_\mathrm{DR,H}}{12}\right)^{9/34}\left(\frac{108}{g_*(T_\mathrm{evap})}\right)^{3/34}\Omega^{-24/17}_\mathrm{PBH,f}.
\eeq
Taking as well into account the Planck upper limit on $\Delta N_\mathrm{eff}$~\cite{Planck:2018vyg}, namely that $\Delta N_\mathrm{eff}\leq 0.3$, one gets an upper bound on the number of the DR degrees of freedom,
\beq
g_\mathrm{DR,H}<12\left(\frac{g_*(T_\mathrm{evap})}{108}\right)^{1/3}.
\eeq

\subsection{$\Delta N_\mathrm{eff}\simeq\Delta N_\mathrm{eff} \rvert _\mathrm{SIGW}$ }
On the other hand,  if the dominant contribution to  $ \Delta N_\mathrm{eff} $ comes from the SIGWs induced by PBH Poisson fluctuations, then one can extract  a lower bound on $m_\mathrm{PBH}$ which should be larger than $10\mathrm{g}$ for PBHs to form after the end of inflation [See \Eq{eq:domain:mPBHf}]. At the end, this lower bound on $m_\mathrm{PBH}$ translates to an upper bound on the PBH abundance at formation, $\Omega_\mathrm{PBH,f}$ which will read as
\beq
\Omega_\mathrm{PBH,f}\leq 5 \times 10^{-3}\left(\frac{g_\mathrm{DR,H}}{12}\right)^{3/16}\left(\frac{108}{g_*(T_\mathrm{evap})}\right)^{1/16}.
\eeq
In addition,  requiring that $\Delta N_\mathrm{eff} \rvert _\mathrm{SIGW}\leq 0.3$ so as to be consistent with Planck~\cite{Planck:2018vyg}, one gets another upper bound on $\Omega_\mathrm{PBH,f}$ which can be recast as follows:
\beq
\Omega_\mathrm{PBH,f}\leq 2\times 10^{-8}\left(\frac{10^9\mathrm{g}}{m_\mathrm{PBH}}\right)^{34/9}.
\eeq
Thus,  at the end one gets that in order for SIGWs from PBH Poisson fluctuations to contribute dominantly to $\Delta N_\mathrm{eff}$ the PBH abundance at formation time should lie within the following range:
\beq
 10^{-15} \sqrt{\frac{g_\mathrm{*,H}(T_\mathrm{BH})}{108}} \frac{10^9\mathrm{g}}{m_\mathrm{PBH}}\leq \Omega_\mathrm{PBH,f}\leq \min\left[5 \times 10^{-3}\left(\frac{g_\mathrm{DR,H}}{12}\right)^{3/16}\left(\frac{108}{g_*(T_\mathrm{evap})}\right)^{1/16},2\times 10^{-8}\left(\frac{10^9\mathrm{g}}{m_\mathrm{PBH}}\right)^{34/9}\right],
\eeq
where the lower bound on $\Omega_\mathrm{PBH,f}$ comes from the necessary condition to have an early PBH-dominated era [See \Eq{eq:domain:OmegaPBHf}].

\section{Conclusions}\label{sec:conclusions}
The Hawking evaporation of ultra-light PBHs  dominating the early Universe before BBN can efficiently alleviate the $H_0$ tension issue.  In particular, the injection to the primordial plasma of dark radiation light degrees of freedom with feeble couplings to the Standard Model (SM) can potentially increase the effective number of extra neutrino species $\Delta N_\mathrm{eff}$ and subsequently the value of the Hubble parameter at early times reconciling it in this way with its late-time value as measured by late-time observational probes. 

Interestingly,  these light PBHs can form a gas of Poisson distributed compact objects which can induce a gravitational-wave (GW) background due to second order gravitational interactions. Thus, by considering the contribution to $\Delta N_\mathrm{eff}$ from the production of the aforementioned GW background we revisited in this work the constraints on the relevant parameters at hand, namely the PBH mass,  $m_\mathrm{PBH}$, the initial PBH abundance at PBH formation time, $\Omega_\mathrm{PBH,f}$ and the number of DR radiation degrees of freedom, $g_\mathrm{DR}$ by accounting at the same time for the relevant upper bounds constraints on $\Delta N_\mathrm{eff}$ from the Planck collaboration. In this way, we gained an insight on the above mentioned PBH domination mechanism for the alleviation of the $H_0$ tension.

At this point, one should point out that the bounds derived above on $m_\mathrm{PBH}$ and $\Omega_\mathrm{PBH,f}$ will be slightly modified in the case of rotating Kerr-like PBHs.  The biggest change in $\Delta N_\mathrm{eff} \rvert _\mathrm{DR}$ is inferred in the case of graviton emission from PBH Hawking evaporation where for the case of a maximally rotating PBH it is found that $\Delta N_\mathrm{eff} \rvert _\mathrm{DR}$ is $90\%$ increased compared to the case of a Schwarzschild PBH. On the other hand, in the case of other scalar, vector and fermionic DR degrees of freedom the relative change of $\Delta N_\mathrm{eff} \rvert _\mathrm{DR}$ compared to the Schwarzschild case is of the order of a few percents.  For more details you can see~\cite{Domenech:2021wkk} as well as~\cite{Bhaumik:2022zdd} which was released soon after our contribution to the Corfu 2022 Summer Institute.

Finally,  let us mention some interesting future research directions which can be further investigated. Interestingly, one can study the effect of realistic extended PBH mass distributions, where PBHs form with different masses and evaporate at different times prolonging in this way the duration of the early PBH-dominated era and potentially increase $\Delta N_\mathrm{eff} $ from both the DR and the GW contribution~\cite{Papanikolaou:2022chm}.  Another research prospect to be explored is the study of the effect of non-Gaussian features of the primordial PBH energy density perturbations on the associated to them GW signal~\cite{Cai:2018dig}. In particular, by using the aforementioned GW portal one can set constraints on primordial non-Gaussianity by studying the GW contribution to $\Delta N_\mathrm{eff}$.  It would be also interesting to investigate the contribution to $\Delta N_\mathrm{eff}$ of the above mentioned GW background within alternative theories of gravity~\cite{Papanikolaou:2021uhe} constraining at the end their parameter space and testing the current gravity paradigm.

\begin{acknowledgments}
T.P. acknowledges financial support from the Foundation for Education
and European Culture in Greece as well as the contribution of the COST Actions
CA18108 ``Quantum Gravity Phenomenology in the multi-messenger approach'' and  CA21136 ``Addressing observational tensions in cosmology with systematics and 
fundamental physics (CosmoVerse)''.

\end{acknowledgments}

\bibliographystyle{JHEP} 
\bibliography{PBH}
.


\end{document}